\documentclass[a4paper, 11pt]{article}

\usepackage[utf8]{inputenc}
\usepackage[T1]{fontenc}

\usepackage{lmodern}

\usepackage{graphics}
\usepackage[margin=3cm]{geometry}
\usepackage{booktabs}
\usepackage{microtype}
\usepackage{color}
\newcommand{\ra}[1]{\renewcommand{\arraystretch}{#1}}

\begin{document}

\title{\textbf{Memory-only selection of dictionary PINs}}
\author{Martin Stanek \\[1ex]
  Department of Computer Science \\ 
  Comenius University \\
  {\small\texttt{stanek@dcs.fmph.uniba.sk}}}
\date{}
\maketitle

\begin{abstract}
We estimate the security of dictionary-based PINs (Personal Identification Numbers) that 
a user selects from his/her memory without any additional aids. The estimates take into 
account the distribution of words in source language. We use established security metrics, 
such as entropy, guesswork, marginal guesswork and marginal success rate. The metrics are 
evaluated for various scenarios -- aimed at improving the security of the produced PINs. 
In general, plain and straightforward construction of memory-only dictionary PINs yields 
unsatisfactory results and more involved methods must be used to produce secure PINs.
\end{abstract}

\section{Introduction}

A PIN is frequently used form of user authentication. The PIN is a fixed-length string of digits, 
usually of length 4, 5 or 6. There are recommendations on how to choose and work with the PINs 
in secure manner, e.g. \cite{PCI,VISA,ISO}. Other proposals try to devise methods for producing
sufficiently secure PIN \cite{JL11,JL13}.
Even though the users are often informed and aware of PIN security, several studies showed that 
the significant portion of the users choose weak, easily guessable PINs \cite{PINstudy,BPA12}.
Weaknesses can also lie in other aspects of using authentication secrets, e.g. partial password/PIN 
verification \cite{AJ13}.

\begin{figure*}[h]
\begin{center}\includegraphics{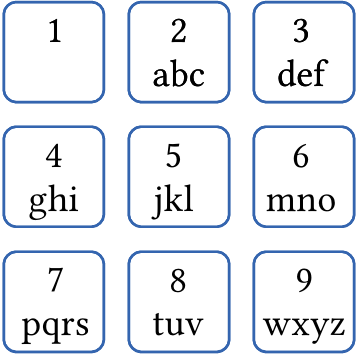}\end{center}
\caption{The standard mapping}
\label{fig-numpad}
\end{figure*}

One possibility of choosing and memorizing the PIN is to use so-called dictionary PIN. Dictionary PIN is
derived from a word with the mapping offered by numpads of ATMs, mobile phones or Point-of-Sale terminals.
The most commonly used letter to digit mapping is the standard mapping shown in Figure \ref{fig-numpad}.

Certainly, other mappings are possible, covering also digits 0 and 1. Recent study of dictionary PINs 
\cite{SS13} analyzed the security of dictionary PINs with respect to various languages and dictionaries.
It also described and assessed several methods of improving dictionary PINs selection. The assessment 
assumed uniform distribution of dictionary words, see \cite{SS13}:

\begin{quote}
``Let us stress that the experiments treat the words in a dictionary as equally probable. This is certainly 
not true if a user chooses the word from his/her memory. The uniform distribution can be easily 
achieved with the aid of an application that offers random sets of words for the user to choose from.''
\end{quote}

Nevertheless, sometimes it can be impractical to use an external application and some users can hesitate
to use or they would not even trust such application for PIN selection. Therefore we focus on dictionary
PINs that user selects from his/her memory, without any external aid. We address the question of the security
of such user-generated dictionary PINs in this paper. The main findings of our experiments are the following:

\begin{itemize}
\renewcommand{\labelitemi}{$-$}
\item Considering uniform frequencies of dictionary words is inadequate for estimating the security of memory-only
  selection of dictionary PINs. 
\item The straightforward word to PIN translation yields unacceptable marginal success rates when frequencies
  are taken into account.
\item Simple blacklisting, prefix, and two-dictionary methods offer only a moderate improvement in security
  metrics.
\item A more demanding methods, such as morphing or combination of multiple methods, are needed to obtain
  significantly better results.
\end{itemize}

\subsection{Quantifying the predictability of PIN}
\label{sec-metrics}

Let $N=10^n$ be the size of an PIN space, for PIN length $n$. Let $X$ be a random variable over the set $\{0,1,\dots,9\}^n$. 
Let $p_i$ denotes the probability of choosing a particular PIN $x_i$. Without loss of generality we assume 
that PINs are sorted in the descending order of their probabilities, i.e.\ $p_1\geq p_2\geq \ldots\geq p_N$.

Various measures for the PIN choices were proposed and studied, for details, discussions and relations 
among these metrics see \cite{BJM10,BPA12,W10}. The most important ones are defined in the following list.

\begin{itemize}
\renewcommand{\labelitemi}{$-$}
\item Shannon entropy, expressed in bits, measures the uncertainty in a random variable: 
  \[ H_1(X) = - \sum_{i=1}^N p_i \log_2 p_i. \]
\item The guesswork measures the expected number of guesses needed to find a PIN, trying in descending 
  order according their probability:
  \[ G(X) = \sum_{i=1}^N i\cdot p_i. \]
\item The marginal guesswork measures the expected number of guesses needed to increase the success 
probability of finding the PIN to at least $\alpha$ (usually $\alpha = 0.5$):
 \[ \mu_\alpha(X) = \min \{1\leq k \leq N\mid {\textstyle\sum\nolimits_{i=1}^k p_i} \geq \alpha\}. \]
\item The marginal success rate measures the probability of guessing the PIN given $\beta$ attempts
  ($\beta$ is usually $3$ or $6$):
  \[ \lambda_\beta = \sum_{i=1}^\beta p_i. \]
\end{itemize}

Since the guesswork and the marginal guesswork are not directly comparable to Shannon entropy, 
the values of $G(X)$ and $\mu_\alpha(X)$ are converted into bits using the following formulas \cite{BJM10}:
$\tilde G(X) = \log_2(2G(X)-1)$, $\tilde \mu_\alpha(X) = \log_2 (\mu_\alpha(X)/\lambda_{\mu_\alpha})$.

Since the standard mapping does not cover digits 0 and 1, the ideal security metrics have the following 
values for dictionary PINs (assuming uniform distribution of PINs) :

\begin{itemize}
\renewcommand{\labelitemi}{$-$}
\item PIN length 4: $H_1(X) = \tilde G(X) = \tilde \mu_{0.5} = 12.00$ bits, $\lambda_6(X)= 0.15\%$,
\item PIN length 5: $H_1(X) = \tilde G(X) = \tilde \mu_{0.5} = 15.00$ bits, $\lambda_6(X)= 0.02\%$.
\end{itemize}

\section{Estimating metrics for dictionary PIN selection}

In order to model frequency distribution of words in a language we use frequency lists based on subtitles.
This is a respected method for analyzing contemporary languages \cite{BN09}. We use two frequency lists for
English -- a carefully prepared SUBTLEXus \cite{SUBTLEXus} containing 60,384 words with a frequency 
higher than 1, and the list compiled from subtitles available from opensubtitles.org \cite{opensub}, 
containing more than 450,000 words (even words with frequency 1). We will denote the results obtained using 
the first/second list with label ``SUBTLEXus''/``opensub'', respectively.

\subsection{Basic statistics}

We compare the metrics for straightforward translation of words to PINs using the standard mapping, see Figure \ref{fig-numpad}, with results obtained in \cite{SS13}. We consider only the words with the 
length equal to the PIN length $n$. The translation starts with stripping the diacritical marks, if they are present. 
Then, the word is mapped to PIN using the standard mapping, e.g. ``love'' and ``hate'' yield 5683 and 4283,
respectively. The frequency of particular word contributes to the probability of resulting PIN.

The results for the PIN lengths 4 and 5 are shown in Table \ref{tab1-comp}. The columns labeled ``uniform''
contain results for PINs derived from a large (spell-checker) English dictionary assuming uniform frequencies 
of words \cite{SS13}. The columns labeled ``RockYou'' contains results for PINs derived from RockYou password 
database where frequencies of words (passwords) were taken into account. It is easy to notice a striking 
difference between scenarios that consider the frequencies of words and those that do not. 

\begin{table*}[h]\centering
\ra{1.2}
\begin{tabular}{@{}lrrrrrrrrrrr@{}}\toprule
 & \multicolumn{2}{c}{SUBTLEXus} &\phantom{a} & \multicolumn{2}{c}{opensub} &\phantom{a}& \multicolumn{2}{c}{uniform \cite{SS13}}
   &\phantom{a}& \multicolumn{2}{c}{RockYou \cite{SS13}}\\
   \cmidrule{2-3} \cmidrule{5-6} \cmidrule{8-9} \cmidrule{11-12}
 & $n=4$ & $n=5$ && $n=4$ & $n=5$ && $n=4$ & $n=5$  && $n=4$ & $n=5$ \\ 
\midrule
$H_1$ (bits) 
  & $7.23$ & $8.42$ && $7.42$ & $8.88$ && $11.28$ & $13.37$ && $-$ & $-$ \\
$\tilde G$  (bits) 
  & $7.18$ & $8.63$ && $7.49$ & $9.45$ && $10.94$ & $13.08$ && $-$ & $-$ \\
$\tilde \mu_{0.5}$ (bits) 
  & $5.52$ & $6.58$ && $5.64$ & $6.92$ && $10.61$ & $12.68$ && $9.20$ & $10.76$ \\
$\lambda_{6}$ (\%) 
  & $23.93$ & $19.24$ && $22.80$ & $17.34$ && $0.85$ & $0.33$ && $10.81$ & $8.46$ \\
\bottomrule
\end{tabular}
\caption{Comparison of metrics for straightforward construction of dictionary PINs}\label{tab1-comp}
\end{table*}

A closer look at the most frequent dictionary PINs from SUBTLEXus and opensub reveals the following observations:

\begin{itemize}
\renewcommand{\labelitemi}{$-$}
\item PIN length 4: Top 7 PINs share the same spots in SUBTLEXus and opensub scenarios (the last three spots in 
top 10 are just permuted, i.e. the top 10 contains exactly the same set of PINs in both scenarios). The high marginal 
success rates are caused by the frequencies of the following PINs: 8428 (probability $6.65\%$ based on SUBTLEXus, e.g.
produced from the word ``that''), 9428 ($4.64\%$, ``what''), 8447 ($3.76\%$, ``this''), 4283 ($3.14\%$, ``have''), 
9687 ($3.04\%$, ``your''), and 5669 ($2.70\%$, ``know'').
\item PIN length 5: The sets of top 10 PINs differ in just two PINs, while the first 5 spots are exactly the same
in SUBTLEXus and opensub scenarios. For SUBTLEXus scenario the most frequent PINs are: 84373 ($5.12\%$, e.g. produced 
from the word ``there''), 74448 ($3.95\%$, ``right''), 22688 ($3.54\%$, ``about''), 84465 ($2.62\%$, ``think''),
46464 ($2.07\%$, ``going''), and 46662 ($1.94\%$, ``gonna'').
\end{itemize}

We can conclude that considering uniform frequencies of dictionary words is inadequate for estimating the security of memory-only
selection of dictionary PIN. The results show the deficiency of such PINs clearly -- the marginal success
rates are unacceptable. Moreover, it seems that this strategy is worse (in average) than strategies currently employed by
users. In order to compare memory-only dictionary PINs with ``common'' PIN selection strategies, we present estimates 
of PIN metrics based on RockYou password database and iPhone unlock codes \cite{BPA12} in Table \ref{tab2}.
On the other hand, the lack of digits 0 and 1 in the standard mapping ensures that these digits do not appear in the 
resulting PIN. Therefore the PINs that are often blacklisted, e.g. 0000, 1111 or 1234, or those the users are warned not to 
use, e.g. birthday or anniversary years, cannot be selected this way.

\begin{table*}[h]\centering
\ra{1.2}
\begin{tabular}{@{}lrrr@{}}\toprule
 & RockYou & iPhone \\
 \midrule
$H_1$ (bits)       & $10.74$ & $11.42$ \\
$\tilde G$  (bits) & $11.50$ & $11.83$ \\
$\tilde \mu_{0.5}$ (bits) & $9.11$ & $10.37$ \\
$\lambda_{6}$ (\%) & $12.29$ & $12.39$ \\
\bottomrule
\end{tabular}
\caption{Estimation of PIN metrics (PIN length 4) \cite{BPA12}}\label{tab2}
\end{table*}

We explore few possibilities for improving memory-only dictionary PINs in the following sections.

\subsection{Blacklisting}

Blacklisting is a common method for improving the security of user-selected PINs. Even if not strictly enforced
(by forbidding the selection of some PINs), at least there are recommendations what PINs a user should not
choose, e.g. see \cite{VISA}:

\begin{quote}
``Select a PIN that cannot be easily guessed (i.e., do not use birth date, 
partial account numbers, sequential numbers like 1234, or repeated 
values such as 1111).''
\end{quote}

\begin{figure*}[h]\centering
\begin{center}\includegraphics{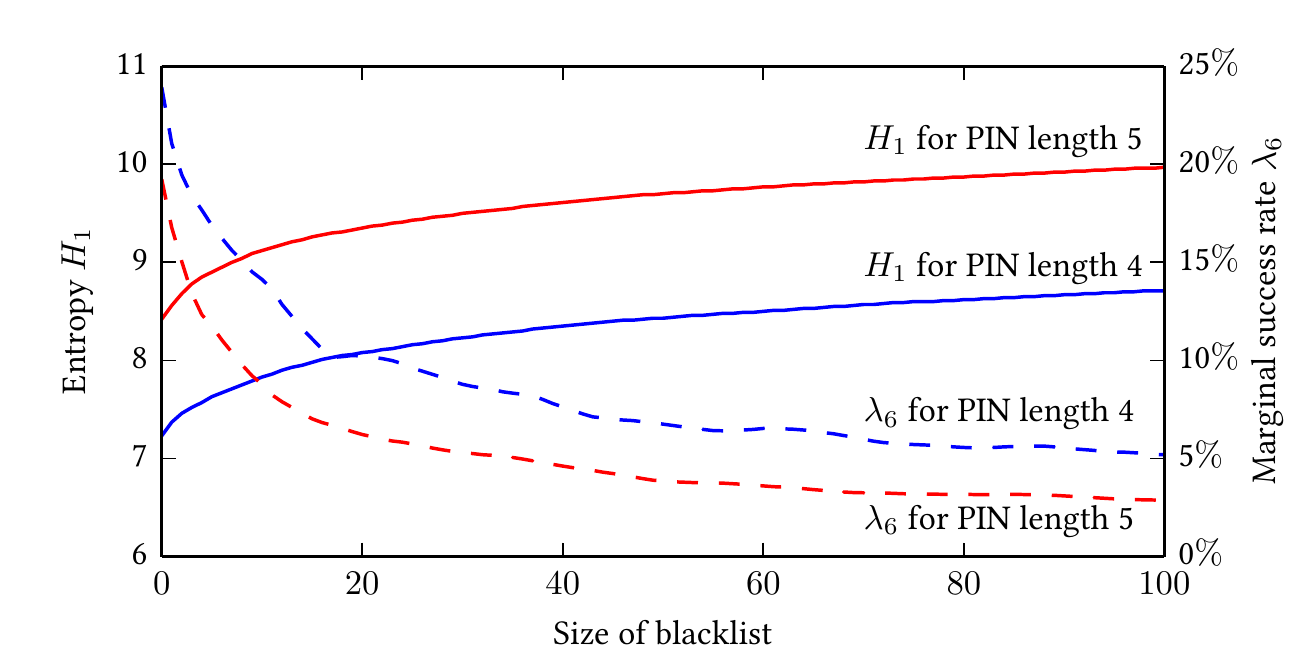}\end{center}
\caption{The effect of PIN blacklisting on entropy and marginal success rate}\label{fig1-black}
\end{figure*}

There are two possibilities for blacklisting in dictionary PIN scenario: first, blacklisting the most frequent 
words; and second, blacklisting the most frequent PINs. The PIN blacklisting is easier to enforce in practice,
and the values of security metrics are comparable for both approaches. Figure \ref{fig1-black} shows the entropy 
and the marginal success rate ($\lambda_6$) for PIN blacklisting (based on SUBTLELXus) ranging from $0$ to $100$
blacklisted PINs. The results show only a moderate improvement in security metrics, therefore blacklisting alone
is not a satisfactory method for improving memory-only dictionary PINs.

\subsection{Modifications of PIN construction}

In order to improve the security metrics of resulting PIN, some natural modifications to basic translation
of dictionary word to PIN were proposed in \cite{SS13}. We evaluate these modifications when applied 
to our ``frequency-aware'' experiments:

\begin{itemize}
\renewcommand{\labelitemi}{$-$}
\item Stretched mapping -- in order to cover digits $0$ and $1$, we can stretch the standard mapping. 
  Our estimates for this modification use the following mapping: a, b $\mapsto 1$; c, d  $\mapsto 2$;
  e, f  $\mapsto 3$; g, h, i  $\mapsto 4$; j, k, l  $\mapsto 5$; m, n  $\mapsto 6$; o, p, q  $\mapsto 7$;
  r, s, t $\mapsto 8$; u, v, w $\mapsto 9$; x, y, z $\mapsto 0$.
\item Prefix -- instead of taking just words with the desired PIN length, i.e. $n$, we use any word with the
  length greater or equal to $n$ and we use its prefix for translation to PIN.
\item Morphing -- the standard word to PIN translation is enriched by random change of one character.
  Assuming that user can choose a random position in a word/PIN and a random digit, the resulting PIN
  is formed by replacing this position by the chosen digit. For example ``this'' can be translated to
  ``t1is''/8147, ``0his''/0447, ``thi9''/8449, etc. Certainly, this method is more demanding than the
  straightforward use of dictionary words. Our estimates assume uniform distribution of positions and
  digits for this method.
\end{itemize}

The results for all above methods are presented in Table \ref{tab3}. The stretched mapping yields no 
improvement at all -- the most frequent PINs changed their values, but their frequencies remained almost
unchanged. The prefix method offers a moderate improvement in security metrics. Obviously, the morphing 
is the most successful approach by a wide margin.

\begin{table*}[h]\centering
\ra{1.2}
\begin{tabular}{@{}lrrrrrrrrr@{}}\toprule
 & \multicolumn{2}{c}{Stretched map.} &\phantom{a} & \multicolumn{2}{c}{Prefix} &\phantom{a} & \multicolumn{2}{c}{Morphing} \\
   \cmidrule{2-3} \cmidrule{5-6} \cmidrule{8-9}
 & $n=4$ & $n=5$ && $n=4$ & $n=5$ && $n=4$ & $n=5$ \\ 
\midrule
$H_1$ (bits) & $7.28$ & $8.43$ && $9.03$ & $10.36$ && $11.08$ & $12.96$ \\
$\tilde G$  (bits)  & $7.28$ & $8.67$ && $8.89$ & $10.39$ && $10.73$ & $12.84$ \\
$\tilde \mu_{0.5}$ (bits) & $5.52$ & $6.58$ && $7.56$ & $8.77$ && $9.88$ & $11.53$ \\
$\lambda_{6}$ (\%)  & $24.04$ & $19.31$ && $11.30$ & $8.01$ && $2.77$ & $2.17$ \\
\bottomrule
\end{tabular}
\caption{Modifications of PIN constructions -- results based on SUBTLEXus}\label{tab3}
\end{table*}

Comparing these results with the estimates from Table \ref{tab2}, we can notice that the prefix method
for dictionary PINs offers slightly better marginal success rate but worse entropy, guesswork and marginal guesswork. 
An interesting observation is that the morphing offers much better marginal success rate while keeping other
security metrics comparable to real-word estimates from Table \ref{tab2}.

Interestingly, the prefix method and the morphing yield better results than the estimates of PIN entropy 
by NIST \cite{NIST}: 9 and 10 bits for PIN lengths 4 and 5, respectively. On the other hand, plain memory-only
dictionary PINs offer less entropy than these estimates.

\subsection{Blacklisting the prefix and the morphing methods}

We expect that blacklisting of the most frequent PINs can further improve the security metrics 
of promising methods from the previous section (i.e. prefix and morphing methods). Indeed, our 
experiments confirm this expectation. Table \ref{tab4} shows the values of the entropy and the
marginal success rate for various sizes of the blacklist ($0$, $10$, and $20$). 

\begin{table*}[h]\centering
\ra{1.2}
\begin{tabular}{@{}lrrrrrrrrr@{}}\toprule
 &&& \multicolumn{2}{c}{Prefix} &\phantom{a} & \multicolumn{2}{c}{Morphing} \\
   \cmidrule{4-5} \cmidrule{7-8}
 & blacklist &\phantom{a} & $n=4$ & $n=5$ && $n=4$ & $n=5$ \\ 
\midrule
$H_1$ (bits) 
  & 0 && $9.03$ & $10.36$ && $11.08$ & $12.96$ \\
  &10 && $9.37$ & $10.68$ && $11.15$ & $13.06$ \\
  &20 && $9.53$ & $10.82$ && $11.19$ & $13.09$ \\
\midrule
$\lambda_{6}$ (\%) 
  &0  && $11.30$ & $8.01$ && $2.77$ & $2.17$ \\
  &10 && $6.62$ & $4.28$ && $1.74$ & $0.97$ \\
  &20 && $4.69$ & $2.95$ && $1.56$ & $0.85$ \\
\bottomrule
\end{tabular}
\caption{Combination of PIN blacklist and the prefix/morphing method}\label{tab4}
\end{table*}

The blacklisting substantially improves the marginal success rate, but offers only a moderate
improvement of the entropy. A disadvantage of the blacklisting is that it complicates the
implementation of authentication.

\subsection{Two-dictionary PINs}

Many people know more than one language. In such case, it is easy to adopt a strategy where a user
randomly choses a language and then (s)he selects a word for PIN construction. We expect obviously
an improvement in security metrics values. In order to assess the improvement we use English 
and Dutch frequency dictionaries SUBTLEXus and SUBTLEXnl \cite{SUBTLEXnl}. We use words with
frequency above 1 in both dictionaries, and we assume that the user selects the dictionary with
probability $1/2$. The results for this two-dictionary scenario are shown 
in Table \ref{tab5}, where ``basic'' denotes the construction using words with the length $n$, 
``prefix'' denotes the prefix method, and ``prefix (BL)'' denotes the combination of the prefix 
method with the blacklist of the length $10$.

\begin{table*}[h]\centering
\ra{1.2}
\begin{tabular}{@{}lrrrrrrrrr@{}}\toprule
 & \multicolumn{2}{c}{Basic} &\phantom{a} & \multicolumn{2}{c}{Prefix} &\phantom{a} & \multicolumn{2}{c}{Prefix (BL 10)} \\
   \cmidrule{2-3} \cmidrule{5-6} \cmidrule{8-9}
 & $n=4$ & $n=5$ && $n=4$ & $n=5$ && $n=4$ & $n=5$ \\ 
\midrule
$H_1$ (bits)              & $7.84$  & $9.35$  && $9.62$ & $11.21$ && $9.84$ & $11.37$ \\
$\tilde G$  (bits)        & $7.72$  & $9.41$  && $9.37$ & $11.09$ && $9.50$ & $11.17$ \\
$\tilde \mu_{0.5}$ (bits) & $6.24$  & $7.63$  && $8.27$ &  $9.68$ && $8.55$ & $9.85$ \\
$\lambda_{6}$ (\%)        & $18.00$ & $11.07$ && $7.78$ &  $4.37$ && $4.09$ & $2.32$ \\
\bottomrule
\end{tabular}
\caption{Security metrics for two-dictionary scenario}\label{tab5}
\end{table*}

As expected, the results are better than results for corresponding single-dictionary scenario methods. 
However, even with the blacklisting the results cannot match the morphing method results for single
dictionary.

\section{Conclusion}

We analyzed the security of memory-only selection of dictionary PINs. The results show that
plain construction of dictionary PINs is unsatisfactory and more involved methods should be used for 
improved security metrics.

\subsection*{Acknowledgment} The author acknowledges support by VEGA 1/0259/13.


\end{document}